# Photonics detection of molecular-specific spatial structural alterations in cell nuclei due to chronic alcoholism and probiotics treatments on colon cancer via a light localization method using confocal imaging

Ishmael Apachigawo[1], Dhruvil Solanki[1], Santanu Maity[1], Pradeep Shukla[2], Radhakrishna Rao[2], Prabhakar Pradhan[1*]

[1] *Department of Physics and Astronomy, Mississippi State University, Mississippi State, MS 39762, USA*
[2] *Department of Physiology, University of Tennessee Health Science Center, Memphis, TN 38163, USA*

Corresponding Author: PP

**Abstract:** Photonics/light localization techniques are important in understanding the structural changes in biological tissues at the nano- to sub-micron scale. It is now known that structural alteration starts at the nanoscale at the beginning of cancer progression. This study examines the molecular-specific nano-structural alterations of chronic alcoholism and probiotic effects on colon cancer using a mouse model of colon cancer. Confocal microscopy and mesoscopic light-scattering analysis are applied to quantify structural changes in DNA (chromatin), cytoskeleton, and ki-67 protein cells with appropriate staining dyes. We assessed alcohol-treated and azoxymethane (AOM) with dextran sulfate sodium (DSS)-induced colitis models, including ethanol (EtOH) and probiotic (*L.Casei*) treatments separately and together. The inverse participation ratio (IPR) technique was employed to quantify the degree of light localization to access the molecular-specific spatial structural disorder as a biomarker for cancer progression detection. Significant enhancement of cancer progression was observed in the alcohol-treated group, and probiotics treatment with alcohol showed partial reversal of these changes in colon cancer. The results underscore the potential of the IPR technique in detecting early structural changes in colon cancer, offering insights into the mitigating effects of probiotics on alcohol-induced enhancement of colon cancer.

## 1. Introduction

Light probing cells/tissues for structural properties and characterization has become relevant in studying cell/tissue abnormalities by understanding the interaction of light waves in disordered media[1–3]. Current research worldwide employs various imaging techniques of light localization to understand and explain tissue-like soft-disordered media[4–8]. Over the years, the importance of light scattering techniques has been highlighted and exploited for many applications. In principle, the nano-to-sub-micron structures in weakly disordered media like cells/tissues change when encountered by light. These changes correspond to the localization among the various nanoscale cell structures and can be seen as the intensity fluctuates for different cells/tissues. In many applications, these scattered light signals through the intensity variations are analyzed statistically and quantified based on the biological sample's changes in the optical parameters (like changes in optical intensities or refractive index). In most biological cells/tissues, these changes in cell/tissue abnormalities only appear in the later stages of the disease. It is known that changes in cells/tissue organs begin early in the nanoscale structure of the cell at the molecular level; however, due to the diffraction limit of light-particle resolution (~200nm), these changes are not apparent in current optical microscopy [9]. Understanding these significant changes for potential biomarkers in early diagnosis and detection methods is essential. This is the ideal motive of this study based on the hypothesis that we can quantify molecular-specific spatial structural alteration by measuring the changes in the physical and intrinsic properties of the cells/tissues. We obtain statistical information for the structural changes by analyzing the reflected light intensity with the mesoscopic wave-scattering framework. Previous works have shown great success in being helpful to quantify nanoscale changes in cells, and there has been great success in its efficacy in distinguishing between different cancer types[3,7,9,10].

Chronic alcoholism has become a global concern regarding the adverse effects it has on organs, particularly the colon. In the US, colon cancer is ranked one of the highest causes of cancer-related mortality[11]. Studies show that alcohol is one of the significant risks of colon cancer, affecting gut microbiota, which leads to inflammation and genetic mutations that advance adenoma to carcinoma [12–14]. Alcohol's first metabolite, acetaldehyde, is a local carcinogen present in humans but not present in mice; however, studies have shown that alcohol consumption has a significant effect on tumor growth in the methyl azoxymethane(MAM)-induced in rats[15,16]. With regards to colon cancer, inflammation in the intestinal mucosal has been reported, with alcohol consumption worsening the effect by activating mast cells [17]. The nuclear structural alterations within the proteins of colon tissues are closely correlated to changes in the morphology of the nanostructural components within the cell/tissue, leading to mass density fluctuations of cellular materials. For example, cell nuclei, DNA, RNA, and proteins have various functions that respond to toxicity by changes in their structural composition, affecting their regular functions. In this case, the effect of alcohol on these proteins is usually affected by the proliferation and rearrangements of macromolecules in the cell[18].

As stated, diseases and abnormalities in cells/tissues like cancer involve intracellular structural changes from the nano to sub-micron scales [1,19]. These changes occur due to the minute rearrangements of the basic building blocks at the molecular level [20]. The nucleus is an important site for our study because it is one of the essential organelles where all cell activities occur[21]. Progress of carcinogenesis, stemming from the nuclei cells and membranes, is altered as carcinogenesis progresses in the nano-submicron level due to intra-molecular re-arrangements. Some nuclear structural alterations occur, leading to spatial mass density fluctuations in the nanoscale for various length scales[20,22]. Standard methods must be more helpful in characterizing optical disordered media due to the spatial heterogeneity from the many kinds of spatial correlation decay length scales within cells/tissues. Quantifying light localization via confocal imaging is significant in quantifying spatial structural disorder. Confocal images are used to construct optical lattices representing the 2D refractive index map corresponding to a particular molecule's spatial mass density distribution. This study's efficacy is achieved using mesoscopic scattering physics-based models to identify and quantify structural disorder *($L_d$)*. Understanding the nanostructural effects of alcohol consumption on a healthy colon tissue enables us in our quantification to confirm the reversing effects of probiotics *(L.Casei)* on alcohol and carcinogenic tissues as reported in a previous study regarding the role of *L.Plantarum* reversible impact on brain cells[23]. Using appropriate confocal imaging with proper probing laser light and treated cells with appropriate dye, we want to track the molecular specific mass density fluctuations to access the normal or abnormal state of the cells via the prominent organelles, including DNA/chromatin, cytoskeleton, and ki67 cells and the treatment of the damages with probiotics like *L.Casei*.

In this study, we calculate the effect of chronic alcoholism on colon cancer as well as the effect of probiotic and chronic alcoholism on colon cancer, using a mice model by quantifying changes in the nanoscale structure of DNA/chromatin, cytoskeletons, etc. We assess and quantify these changes by analyzing the confocal images of colon tissues for different types of proteins, namely, chromatin, cytoskeleton, and ki-67 cells, using molecular-specific fluorescent dyes. The confocal images are taken for normal/wild type mice, which serve as the control, azoxymethane (AOM) with dextran sulfate sodium (DSS) treated colon cancer mice with and without chronic alcoholism. Different cell nuclei were accessed (i) Control/or pair-fed (PF) (ii) Alcohol(ethanol) treated mice(EF) (iii) azoxymethane (AOM) treated + DSS that induces colitis (AOM + DSS); (iv) AOM + DSS combined with ethanol (AOM + DSS + EtOH), and (v) Probiotics treatment with alcohol in AOM/DSS model (AOM + DSS + EtOH + *L.Casei*). We measure the degree of the disorder by calculating the structural disorder $L_{d\text{-}IPR}$ using the IPR technique described elsewhere. [24,25].

## 2. Methods

*Mathematical Model*

As stated, the theoretical basis for our quantification has been fully described elsewhere [23,26–28], but we present a brief discussion of the IPR method for the completeness of this paper. The model for quantifying the disorder from our confocal micrographs is primarily based on finding the eigenvalues and eigenfunctions of optical lattice using the Anderson Tight Binding Model (TBM) [29–31]. In principle, the confocal micrographs are analyzed by a numerical algorithm, where the images are represented as a matrix of various pixel intensity values. Each intensity pixel represented by a specific value corresponds to the molecular-specific mass density fluctuations in a small volume attached to the staining dye due to the refractive index fluctuations in the media. The pixel intensities are used to construct an optical lattice where electron localization is measured using the Inverse Participation ratio (IPR).

The IPR quantifies structural alterations from nano to sub-micron scales in biological cells/tissues [32]. As reported, light localization is assessed through the optical lattice's properties from the pixel intensities of a confocal micrograph. This light localization is quantified by measuring the degree of structural alterations due to the molecular spatial density distribution. It has been established in the literature[2,18] that the structural disorder is proportional to the system's mass density fluctuations or refractive index fluctuation in a weakly disordered system, previous work details the mathematical description of this relation with the final equation written below:

$$I(x,y) \propto \rho(x,y) \propto n(x,y) \qquad (1)$$

$$I_0 + dI(x,y) \propto \rho_0 + d\rho(x,y) \propto n_0 + dn(x,y) \qquad (2)$$

*dn (x, y)* and *dI (x,y)* are the fluctuation part of the refractive index and intensity, respectively.

This relationship allows us to calculate the individual intensity values from the confocal image to reflect the spatial density distribution from the optical lattice with its potential. The optical potential can be shown as: $(\varepsilon_i(x,y))$.

$$\varepsilon_i(x,y) = \frac{dn(x,y)}{n_o} \propto \frac{dI(x,y)}{I_0} \qquad (3)$$

The figure below details the steps for creating the optical lattices whose potentials are calculated using the equation above.

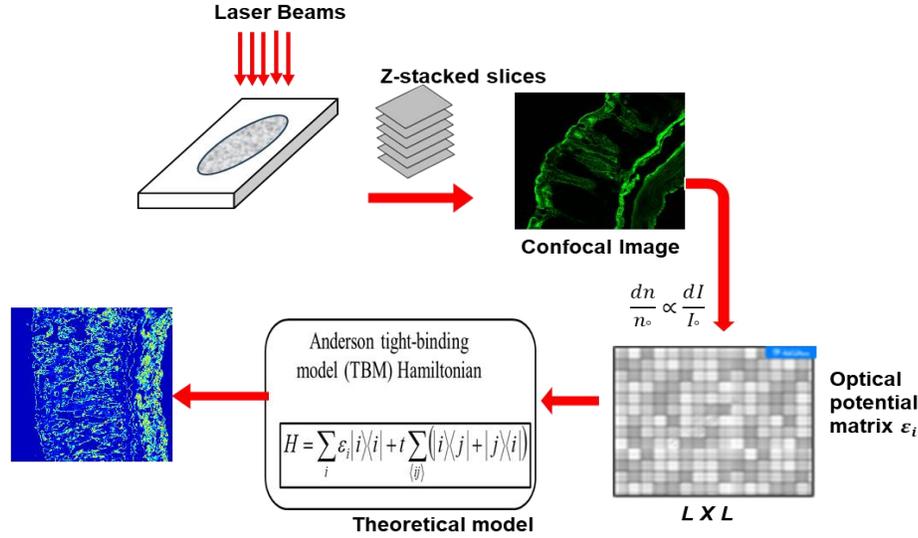

**Fig 1**: **Construction of a disordered optical lattice from confocal imaging (schematic pictures)**. An image is created by creating an exciting sample with a laser beam and scanning each voxel to construct a 2D confocal image of a phalloidin-stained cytoskeleton. The optical potential lattice is created with each pixel value representing the intensity from each voxel of the confocal image.

With the obtained optical potential, we apply the TBM to solve the Hamiltonian of the lattice. This model, as described, allows us to tell a single-optical state of a system in any geometry by defining a Hamiltonian for the system to find the eigenfunctions given as;

$$H = \sum \varepsilon_i |i\rangle\langle i| + t \sum_{\langle ij \rangle} (|i\rangle\langle j| + |j\rangle\langle i|) \qquad (4)$$

where $|i\rangle$ and $|j\rangle$ are the optical wave function or the state of a photon at $i$-th and $j$-th lattice sites, $\langle ij \rangle$ represents the nearest neighbors, $t$ is the interacting parameter between the sites i and j, $\varepsilon_i(x,y)$ is the $i$th lattice site optical potential energy corresponding to the pixel position $(x,y)$. The Hamiltonian's eigenfunctions are computed to facilitate an investigation into the localization of light properties in the disordered optical lattice structure and, as a result, the strengths of the disorder in terms of IPR parameter can be analyzed by measuring how much of a system's total wavefunction is localized in a particular region, and is defined accordingly as

$$\text{IPR} = \int |E(\text{r})|^4 dr \qquad (5)$$

For 2D lattice L×L, the average value of IPR, $\langle IPR \rangle$ over the N eigenfunctions can be computed as;

$$\langle IPR(L)\rangle_{L\times L} = \frac{1}{N}\sum_{i=1}^{N}\int_0^L\int_0^L E_i^4(x,y)dxdy \propto \mathrm{dn}\times l_c \qquad (6)$$

$$\langle IPR(L)\rangle_{L\times L} \sim \mathrm{dn}\times l_c \qquad (7)$$

where $E_i$ is the $i$th eigenfunction of Hamiltonian quantified optical lattices for a small area $L\times L$, and N is the total number of potential points on optical lattices (i.e., $N=(L/dx)^2, dx=dy$). The IPR value indicates how much light is confined or localized in a small area $L\times L$ of a sample, which measures the degree of structural disorder in that area. A higher $\langle IPR(L)\rangle_{L\times L}$ value means that the light is more strongly confined or localized in a closed area, indicating a higher degree of disorder or fluctuations in the sample's refractive index inside the area. The ensemble-averaged IPR value, denoted as $\langle IPR(L)\rangle_{L\times L}$ is acquired by calculating the mean of all the IPR values computed for the entire sample (micrograph) at length scale $L$.

*Sample Preparation*

**Mice models of colon cancer**: For this study, we examine the effects of alcohol and probiotics on precancerous colon tissues, which are essential organs in the digestive system. However, colon cancer is the most prevalent among gastrointestinal cancers. Our sample choices were based on five (5) different mice groups of colon cancer models. A typical characterized model for colon cancer is the AOM + DSS mice model. AOM (azoxymethane), a carcinogen, is used to induce colon cancer in mice in addition to dextran sodium sulfate(DSS), which is a chemical colitogenic that induces colitis in the colon and ensures that there is a 100% probability of developing colon cancer. As part of the study, alcohol (EtOH) and probiotics *like L.Casei* (LC) were treated with carcinogenic mice to study the effect of EtOH and LC on colon cancer tissues. The protocols for colonic tumorigenesis have been reported in earlier works, but for completeness, we present a brief description [18,33]. The mice for the experiments were approved by the Institutional Animal Care and Use Committee (IACUC) of the University of Tennessee Health Science Center (UTHSC), Memphis, TN, following federal and institutional guidelines. The study used 12-14 weeks-old adult mice (C57BL/6, Harlan Laboratories, Houston, TX). They were divided into 5 groups of 4-6 mice per group. These groups were fed a regular Lieber–DeCarli (LDC) liquid diet with or without ethanol (4%, v/v). The control group (PF) and the cancer group (AOM +DSS) had their ethanol diet replaced by isocaloric maltodextrin in their LDC diet for 15/15 days recovery periods. In the first phase of the treatment, colitis was induced in the mice after 5 days of AOM treatment by administering DSS (3% w/v) in five days with a 15-day recovery period. The second phase involved administering DSS again for 5 days for another 15 days recovery period. Ethanol (4%, v/v) was administered in the regular LDC diet during the two cycles of the 15-day recovery periods for the AOM+DSS+EtOH group. For six days, the AOM+DSS+EtOH group was fed with 6% of LC (strain 256, 106 cfu/ml) after 15 days of recovery. On day 45, mice were euthanized and the distal colon tissues were collected from each of the five groups (6 mice for each group): (i) control (PF), (ii) control treated with ethanol (EF), (iii) carcinogenic treated mice with AOM and DSS (AD), (iv) carcinogenic treated mice with EtOH (ADE) and (v) carcinogenic treated mice with EtOH and probiotics (ADE+LC). The tissues were stained with 3 different dyes for specific parts of the nuclear material; namely, DNA/Chromatin was DAPI-stained, Cytoskeleton(F-actin) was stained with Phalloidin, and ki-67 cells were stained with H3K27me3, these were processed for confocal imaging and analysis.

**Confocal Images**

Biopsied colon tissue cryosections (~10μm thick) were fixed in acetone: methanol (1:1) at −20°C for 2 min, and the sections were rehydrated in PBS (*137mM* sodium chloride, *2.7mM* potassium chloride, *10mM* disodium hydrogen phosphate, and *1.8mM* potassium dihydrogen phosphate). The tissue was incubated for 1 hour with secondary antibodies and different types of fluorescent dye. The fluorescence was examined using a Zeiss 710 confocal microscope (Carl Zeiss GmbH, Jena, Germany), and images from *x–y* sections (*1μm*) were collected by LSM 5 Pascal software (Carl Zeiss Microscopy). ImageJ software stacked images (Image Processing and Analysis in Java; National Institutes of Health, USA) and processed them. All the images of tissues from the different groups were collected and processed under identical conditions.

**3. Results**

In this study, we discuss the effects of alcohol and probiotics on colon cancer tissues, precisely three different components in the tissue with different staining dyes (DNA/chromatin, cytoskeleton, Ki-67 cells). We quantified the disorder in the cells within the tissues using light localization techniques by comparing five groups of mice [34]. Our primary biomarker was to probe the confocal micrographs using the IPR technique to evaluate and quantify the nanoscale mass density fluctuations as structural disorder strength $L_{d\text{-}IPR}$. The micrographs were analyzed for disorder strength for length scales at $L = 0.8$ μm for each cell nucleus. The five groups of mice were considered to have no less than 25 micrographs for each category: DNA/Chromatin, Cytoskeleton, and Ki-67 cells. The computed *mean* ⟨IPR⟩ and *std* ⟨IPR⟩ for these categories were compared for the following groups: (i) PF, (ii) EF, (iii) AD, (iv) ADE, and (v) ADE+LC.

The figures below show the intensity variations for the different stained tissues; the intensity variations are dependent on spatial mass density fluctuations within the nuclei; with a higher mass density fluctuation of the attached macromolecules to the dye, there is higher fluctuation in the pixel intensity, which are shown as yellow-red color noise in the images. The statistical significance is represented in the bar graphs below. The significant increase in average $L_d$ and std values for all 3 types of stained tissues shows the effect of EtOH on the colon when EF and PF are compared. Similarly, EtOH affects cancer tissues because it introduces a lot of disorders that cause rearrangement in macromolecules. However, significantly, in all cases studied, it can be seen how effective probiotic LC was in mitigating the effects of EtOH on the carcinogenic mice model, which is shown in the reduction of the *mean* and *std* ⟨IPR⟩ values almost back to normal or less. Figures for the results are shown below.

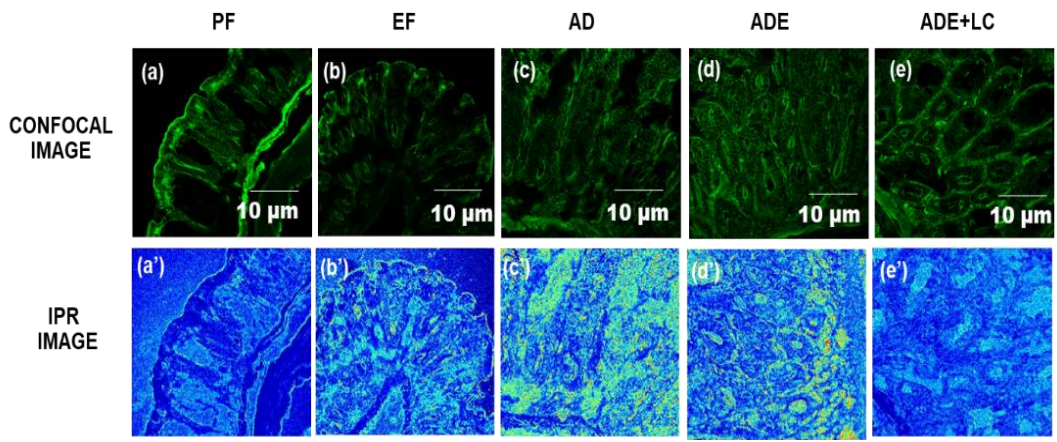

**Figure. 2 Cytoskeleton(F-actin) fibers: Confocal and IPR images of cytoskeleton(F-actin) tissues (PF, EF, AD, ADE, ADE+LC). (a)-(e)** represent the confocal images of the control (PF), Ethanol Fed (EF), Azoxymethne+dextran+sodium sulfate (AD), Azyoxymethene +dextran +sodium sulfate + EtOH (ADE), *L.Casei* (ADE+LC) (Probiotics) while (a')-(e') represent their respective IPR images. In the IPR images, EtOH's structural disorder and influence increase compared to the control (PF) but decrease when probiotics are introduced. This reduces the disorder and the influence of EtOH on the cancer model (AD).

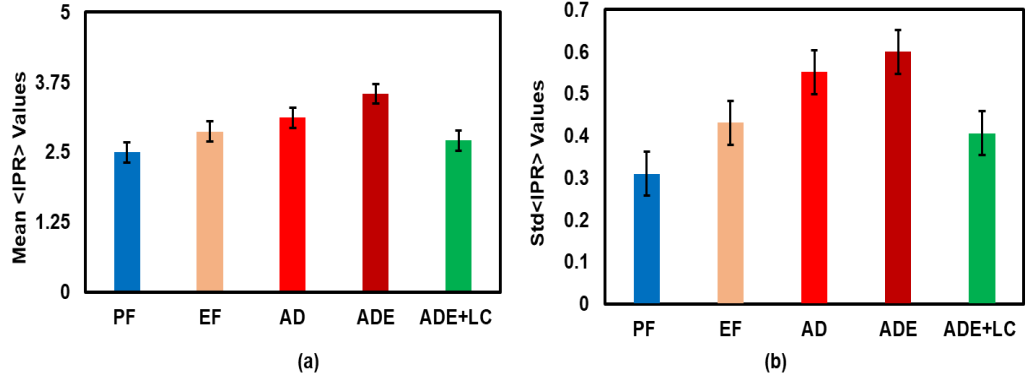

**Fig. 3 Cytoskeleton (F-actin) fibers:** Bar graph representation of the relative values of molecular-specific (a) mean ⟨IPR⟩ and (b) std ⟨IPR⟩ of the cytoskeleton in colon tissue for control (PF) compared to EtOH-fed (EF), Azoxymethane+dextran+sodium sulfate (AD), Azoxymethane+dextran+sodium sulfate +EtOH (ADE) and probiotic fed (ADE+LC). The IPR analysis compares the values of the mean and std for the disorder strength. The chart shows an increase of 15 % and 39% in the structural disorder for the mean and std ⟨IPR⟩ values. This introduces a lot of disorder in the molecular structures of the tissue. Comparatively, AD compared to PF with mean and std values (3.12 and 2.50, respectively) increases by 24% for mean ⟨IPR⟩ and 78% for std ⟨IPR⟩. In the presence of EtOH (ADE), there is a 13% increase in mean ⟨IPR⟩ disorder. However, probiotics take care of the mitigating effects, reducing the structural disorder in the mean ⟨IPR⟩ from ADE by 23% and 32% from std ⟨IPR⟩. (*Student's t-test p-values < 0.05*)

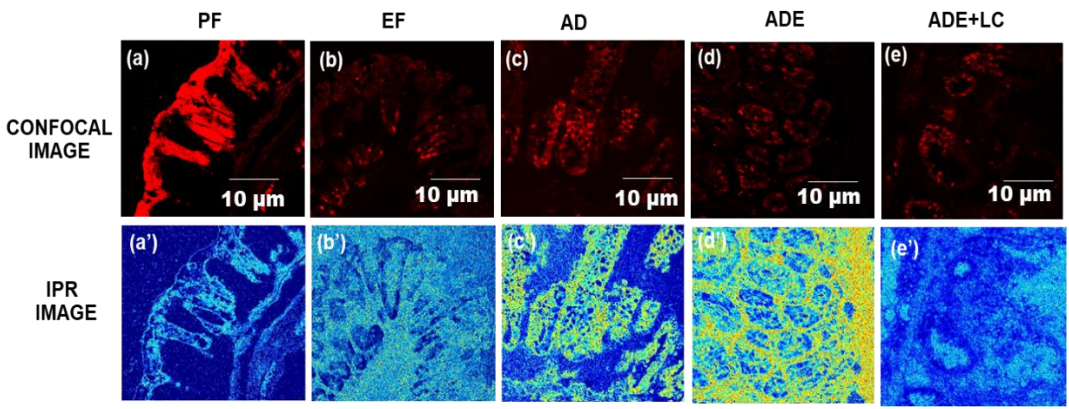

**Fig. 4 Ki-67 cells: Confocal and IPR images of Ki-67 protein cells (PF, EF, AD, ADE, ADE+LC).** (a)-(e) represent the confocal images of the control (PF), Ethanol Fed (EF), Azoxymethane+dextran+sodium sulfate (AD), Azyoxymethane+dextran+sodium sulfate + EtOH (ADE), *L.Casei* (ADE+LC) (Probiotics) while (a')-(e') represent their respective IPR images. In the IPR images, the structural disorder and the influence of EtOH increase compared to the control (PF) but decrease when probiotics are introduced. This reduces the disorder and the influence of EtOH on the cancer model (ADE).

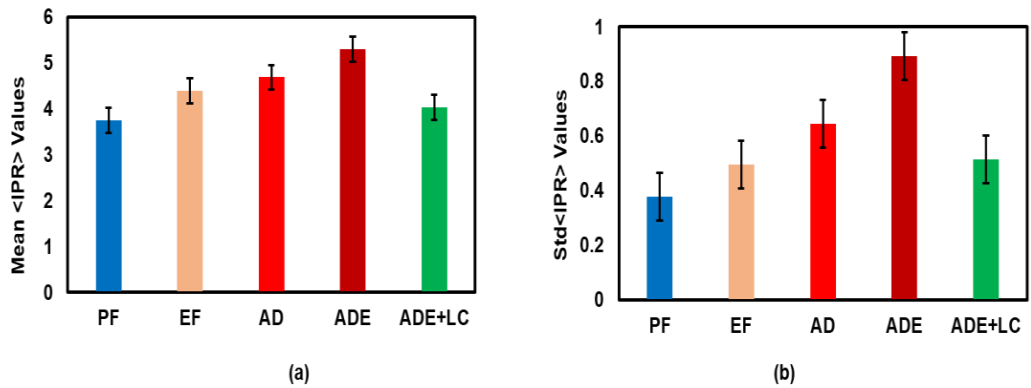

**Fig. 5 Ki-67 cells:** Bar graph representation of the relative values of molecular-specific (a) mean ⟨*IPR*⟩ and (b) std ⟨*IPR*⟩ of the (Ki-67 cells) in colon tissue for control (PF) compared to EtOH-fed (EF), Azoxymethane+dextran+sodium sulfate (AD), Azoxymethane+dextran+sodium sulfate + EtOH (ADE) and probiotic fed (ADE+LC). The IPR analysis compares the values of the mean and std for the disorder strength. From the chart, there is an increase of 17 % and 31% in the structural disorder for the mean and std ⟨*IPR*⟩ values. This introduces a lot of disorder in the molecular structures of the tissue. Comparatively, AD compared to PF with mean and std values (4.69 and 3.74), respectively) increases by 24% for the mean ⟨*IPR*⟩ and 78% for std ⟨*IPR*⟩. In the presence of EtOH (ADE), there is a 13% increase in mean ⟨*IPR*⟩ and a 39% increase in std ⟨*IPR*⟩ disorder. However, probiotics take care of the mitigating effects, reducing the structural disorder in the mean ⟨*IPR*⟩ from ADE by 24% and 42% from std ⟨*IPR*⟩. (*Student's t-test p-values < 0.05*)

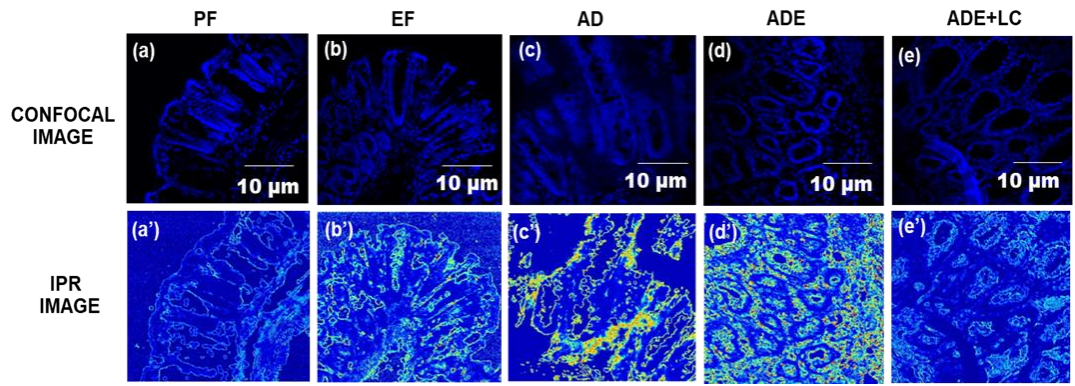

**Fig. 6 DNA/Chromatin: Confocal and IPR images of DNA/chromatin tissues (PF, EF, AD, ADE, ADE+LC).** (a)-(e) represent the confocal images of the control (PF), Ethanol Fed (EF), Azoxymethane+dextran+sodium sulfate (AD), Azyoxymethane+dextran+sodium sulfate + EtOH (ADE), *L.Casei* (ADE+LC) (Probiotics) while (a')-(e') represent their respective IPR images. In the IPR images, EtOH's structural disorder and influence increase compared to the control (PF) but decrease when probiotics are introduced. This reduces the disorder and the influence of EtOH on the cancer model (AD).

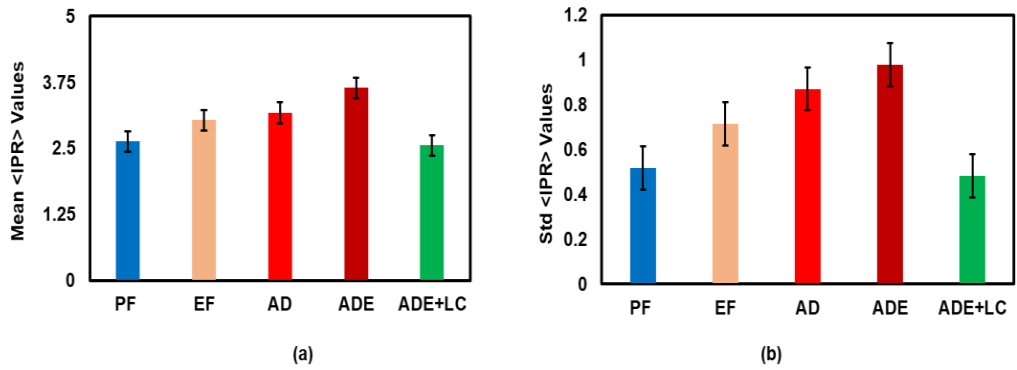

**Fig. 7 DNA/Chromatin:** Bar graph representation of the relative values of DNA molecular-specific (a) mean ⟨IPR⟩ and (b) std ⟨IPR⟩ of DNA/chromatin nuclei in colon tissue for control (PF) compared to EtOH-fed (EF), Azoxymethane+dextran+sodium sulfate (AD), Azoxymethane+dextran+sodium sulfate +EtOH (ADE), and probiotic fed (ADE+LC). The IPR analysis compares the values of the mean and std for the disorder strength. From the chart, there is an increase of 15 % and 39% in the structural disorder for the mean and std ⟨IPR⟩ values. This introduces a lot of disorder in the molecular structures of the tissue. Comparatively, AD compared to PF with mean and std values (3.17 and 2.63), respectively) increases by 21% for mean ⟨IPR⟩ and 68% for std ⟨IPR⟩. In the presence of EtOH (ADE), there is a 15% increase in mean ⟨IPR⟩ and a 12% increase in std ⟨IPR⟩ disorder. However, probiotics take care of the mitigating effects, reducing the structural disorder in the mean ⟨IPR⟩ from ADE by 30% and 51% from std ⟨IPR⟩. *(Student's t-test p-values < 0.05)*

From Figs. 2 and 3, the relative study of the ⟨IPR⟩ values from the *mean* and *std* show the structural and statistical change in the cytoskeleton (F-actin) fibers. There are subcellular changes when the mice diet is replaced with EtOH which is evident in the IPR image in Fig. 2. The relevant change is represented as 15% for the mean $L_{d\text{-}IPR}$ and 39% in terms of the std $L_{d\text{-}IPR}$ fluctuation. EtOH causes a relative increase in the structural disorder by increasing the mass density fluctuations within the cell nuclei. Its influence on the cancer model (ADE) has the highest $L_{d\text{-}IPR}$ value(3.54) and an std value of (0.60) which is a 15% and 14%he bar graph shows a 15% from the $L_{d\text{-}IPR}$ value for PF(2.50) and AD(3.12) respectively; this implies that alcohol harms the colon tissues, particularly the cytoskeleton, which provides structural support to cells and helps understand cell/tissue morphology. This alcohol effect destabilizes the internal organization of the tissue structure, which can be seen in the confocal image. The effects are mitigated by introducing probiotics (ADE+LC), reducing the change in the disorder strength by 23% and 32% for the mean and std $L_{d\text{-}IPR}$ value and correcting the change in cell structure.

Figs. 4 and 5 show the molecular-specific light localization analysis for ki-67 protein cells. The cells are primarily responsible for proliferation assessment in the tissue, where cells are actively divided to enable us to determine the growth fraction of a tumor [35,36]. The $L_{d\text{-}IPR}$ analysis comparing the EtOH-fed (EF) mice show a relative increase in mean and std $L_{d\text{-}IPR}$ of 17 % and 31% compared to the control (PF). The effect is visible in the IPR image, where there is high intensity in fluctuations in the colon tissue. The cancer model (AD) with a structural disorder value of 4.69 of the colon tissue shows a significant change in the disorder strength to 25% and 71% for the mean and std $L_{d\text{-}IPR}$ values when compared to their control (3.75). This cancer model in the presence of EtOH (ADE) has the highest $L_{d\text{-}IPR}$ value and fluctuations (5.30 and 0.89) relative to both PF/EF and AD. This significant increase is a possible reason to explain the division of cells by a percentage increase of 13% in the $L_{d\text{-}IPR}$. The presence of EtOH in cancer-related tissue harms cell division and destroys the structure of the cells, as seen from the IPR images having more yellow-red color noise. Probiotics effectively reduce the impact and cell division to about 24% and 42% in the mean and std $L_{d\text{-}IPR}$ and evident in the IPR images, significantly reducing the yellow-red color higher density fluctuations. This means probiotics are good for lowering the alcohol effect of colon cancer.

Figs. 6 and 7 show the <IPR> images and the bar graph representations for the change in disorder strength of DNA/chromatin, important macromolecules in cell nuclei. The IPR analysis shows that the average value of the disorder strength changes slightly by 15% in the nuclei tissues for EtOH-fed (EF) mice compared with the control (PF). This shows the effect of alcohol on the chromatin of the nuclei cells. However, the relative disorder strength in the std $L_{d\text{-}IPR}$ value shows a much more significant change between EF(3.03) and the PF(2.63) of about 39%. With the cancer model (AD), we see a substantial changes of 21% and 68% increase in mean and std $L_{d\text{-}IPR}$ values, whereas the cancer model in the presence of EtOH shows a change of 15% compared to AD. These changes in the mass density fluctuations of chromatin due to alcohol and its effect on cancer are signs of persistent alterations to the structure of the chromatin, which is responsible for providing information about the cell density and cell functions in the colon. The $L_{d\text{-}IPR}$ value in the presence of probiotics (ADE+LC) shows a sound reduction in the disorder strength for ADE by 30% and 51% in the mean and std fluctuations respectively.

This molecular-specific technique for structural disorder quantification of EtOH and LC on colon tissues, which have been analyzed using the confocal and <IPR> images, shows the linear relationship among all stained tissues in terms of the effect of EtOH on the non-carcinogenic colon and cancer-related colon. This effective disorder is rectified and reversed by probiotics like *L.Casei,* signifying LC-blocking EtOH and carcinogenic-induced changes in the Cytoskeleton, Ki-67 cells, and DNA/Chromatin, mainly reversing the cancer growth.

To summarize the above results, we have drawn summary flow charts for the above results as follows. In the chart below, we compare the $L_{d\text{-}IPR}$ values for the different parts of the cell nuclei( cytoskeleton, DNA/chromatin, Ki-67 cells) by looking at their consistency in terms of the increasing changes in density fluctuations.

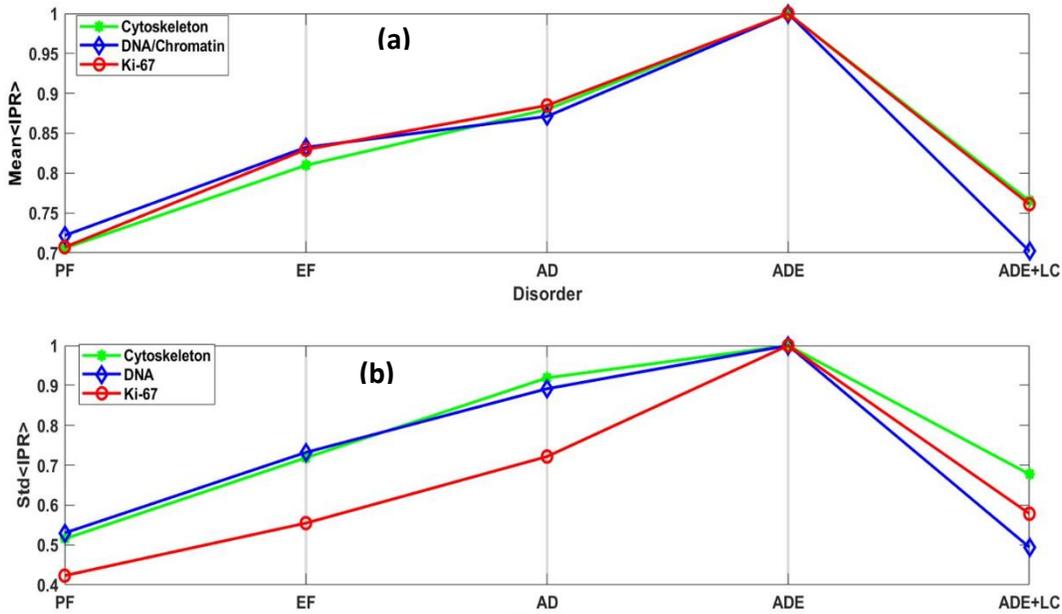

**Fig. 8 Summary flow chart:** The representative flow chart comparing the normalized mean and std ⟨*IPR*⟩ values for various groups at the highest value of the $L_{d\text{-}IPR}$ for (a) Mean IPR and (b) std(IPR). The y-axis shows the increasing mean and std ⟨*IPR*⟩ values and the x-axis represent the various groups (PF, EF, AD, ADE and ADE+LC ). The x-axis steps are false steps that are at an equal distance of unit 1 for each group. The plots show systematic increase in ⟨*IPR*⟩ values except for ADE+LC when probiotics are introduced although each cell nuclei takes a slightly different path. Standard errors are the same as shown in bar graphs. In summary, the cancerous growth increase with the chronic alcoholism, but reduces when treated with probiotics.

## 4. Discussion and Conclusion

We report on the effect of chronic alcoholism and probiotic treatment on normal and cancer cells to study their subcellular/macromolecular structural alterations as a biomarker. We analyzed these structural changes using confocal imaging, which occur at the nanoscale preceding the carcinogenic changes on the microscopic scale. We apply the IPR technique by studying the localization effect of light, which allows us to find the eigenfunctions from the optical lattice structure of the confocal micrograph. These morphological alterations in the nanoscale help us to quantify the structural disorder by a single parameter known as the $L_{d\text{-}IPR}$, which is $L_{d\text{-}IPR} \sim <<IPR(L)>_{LxL} \propto dn^* l_c$, where dn is the refractive index fluctuation, and $l_c$ is the correlation length. Normal mice colon tissues are treated with ethanol; we see an increase in disorder due to the mass density fluctuation due to the structural changes apparent in the tissues for all 3 types of macromolecules/cells (Chromatin, Cytoskeleton, Ki-67 cells). The structural disorder ($L_d$) is increased when compared to the control, which signifies an effect of Ethanol on colon tissues and nuclei. Next, the tissues of the mice are treated with AOM +DSS as the cancer model. We studied the structural change for this model without ethanol. Compared to the control, the $L_{d-IPR}$ value increases significantly for all 3 types of proteins/dyes. The changes are not very apparent in the confocal image but can be seen to change with intensity in the IPR images, as discussed in the results. The progressive carcinogenic changes in the nanoscale structural disorder due to AOM + DSS+ EtOH show the biggest change in disorder, signifying a more rapid change in the mass density fluctuation of the stained tissues or more cancerous growth. The $L_{d-IPR}$ values increase higher compared to other types of tissues discussed above. This means that alcohol enhances the carcinogenic process during early carcinogenesis or enhances the carcinogenic process. The final step is to discuss how the treated carcinogenic mice colon tissue in the presence of alcohol will react with probiotics like *L.Casei*. From all types of protein-stained tissues, i.e., the chromatin, cytoskeleton, and ki-67 cells, respectively, there is a significant reduction in the disorder strength shown by its structural changes, which emanate from the rearrangement of macromolecules, with the progressive carcinogenesis. The decrease in $L_{d-IPR}$ is statistically significant in terms of identifying potential biomarkers.

In effect, chronic alcoholism significantly impacts cancer tissues, particularly in progressive carcinogenesis, by enhancing the tumorigenicity process. However, the results show reversibility to normality in probiotics treatment, where colonic cells try to be more normal. This technique shows potential importance, and $L_{d-IPR}$ could serve as a potential biomarker in early cancer diagnostics and treatments.

**Author Contributions:** PP conceptualized the project; PP and PKS and RKR designed the experiments; IA, DS and SM performed the data analyses, IA and PP wrote the first draft, all authors contributed to final draft.

**Funding:** Part of this work was partially supported by the National Institutes of Health (NIH) grants (R21CA260147) to PP.

**Institutional Review Board Statement:** The animal study received approval from the University of Tennessee Health Science Center, Memphis, TN (IACUC ID 19-010.0-A, Renewal date: 03/2020). All the standard ethical procedures were followed during the study, including personal protection and safety procedures while handling human tissue.

**Data Availability Statement:** Data may be available to the corresponding author, PP, upon request.

**Acknowledgments:** We thank NIH for financial support. We also acknowledge the imaging centers of Mississippi State and UTHSC for confocal imaging.

**Conflicts of Interest:** The authors declare no conflicts of interest.